\documentclass[11pt]{article}

\usepackage{epsfig}

\def\lsim{\mathrel{\rlap{\lower4pt\hbox{\hskip1pt$\sim$}}
    \raise1pt\hbox{$<$}}}         
\def\gsim{\mathrel{\rlap{\lower4pt\hbox{\hskip1pt$\sim$}}
    \raise1pt\hbox{$>$}}}         

\def\overleftrightarrow#1{\vbox{\ialign{##\crcr
    $\leftrightarrow$\crcr
    \noalign{\kern 1pt\nointerlineskip}
    $\hfil\displaystyle{#1}\hfil$\crcr}}}

\def \gta {\mathrel{\vcenter {\hbox{$>$}\nointerlineskip\hbox{$\sim$}}}}

\newcommand{\pp}{\\[0.5cm]}
\newcommand{\be}{\begin{equation}}
\newcommand{\ee}{\end{equation}}
\newcommand{\bea}{\begin{eqnarray}}
\newcommand{\eea}{\end{eqnarray}}

\begin{document}

\hfill {\bf TUM/T39-01-11} \\
\vspace{0.01in}
\hfill {\bf ECT$^*$-01-14} \\
\vspace{0.01in}
\hfill {May 2001} \\

\vspace{0.25in}

\begin{center}
{\bf \Large On the Quasiparticle Description of \\ Lattice QCD
Thermodynamics\footnote{Work
supported in part by BMBF and GSI.}}
\end{center}
\vspace{0.25 in}
\begin{center}
{R.A. Schneider$^{a}$ and  W. Weise$^{a,b}$\\}
\vspace{0.25 in}
{\small \em $^{a}$ Physik Department, Technische Universit\"{a}t M\"{u}nchen, D-85747 Garching,
GERMANY\\}

\vspace{0.05 in}

{\small \em $^{b}$ ECT$^*$,  I-38050 Villazzano (Trento), ITALY}

%

\end{center}
\vspace{0.5 in}

\begin{abstract}
We propose a novel quasiparticle interpretation of the equation of state of deconfined QCD
at finite temperature. Using appropriate thermal masses, we introduce a phenomenological parametrization of the
onset of confinement in the vicinity of the predicted phase transition. Lattice results of the
energy density, the pressure and the interaction measure of pure $SU(3)$ gauge theory are excellently
reproduced. We find a relationship between the thermal energy density  of the Yang-Mills
vacuum
and the chromomagnetic condensate $\langle \mathbf{B}^2 \rangle_T$. Finally, an extension to QCD with
dynamical quarks is discussed. Good agreement
with lattice data for 2, 2+1 and 3 flavour QCD is obtained. We also present the QCD
equation of state
for realistic quark
masses.
\end{abstract}
\newpage
\section{Introduction}
There are convincing arguments that QCD exhibits a transition from a confined hadronic phase to a
deconfined partonic phase, the quark-gluon plasma (QGP), at a temperature of
$T_C \sim 170$ MeV (for two light quark flavours) \cite{AK01, FKE00}. A central quantity of matter in
thermal equilibrium is the Helmholtz free energy from which the pressure $p$, energy density
$\epsilon$ and entropy density $s$ can be derived. These entities are important for the description of,
{\em e.g.}, ultra-relativistic heavy-ion collisions at the CERN SPS and RHIC, or the evolution of the
universe after the first $10^{-6} - 10^{-5}$ seconds. The challenge lies therefore in the
derivation of the equation of state (EoS) of hot, deconfined QCD from first
principles. Perturbative results are available
up to order $\mathcal{O}(g_s^5)$ \cite{ZK95}. However, for temperatures of interest in
the experimentally accessible region (a few
times
$T_C$), the strong coupling constant is presumably large: $g_s \simeq 1.5 - 2$. The perturbative
expansion in powers of $g_s$  shows
bad convergence already for much smaller values of the coupling. Furthermore, in the vicinity of a phase
transition, perturbative methods are in general not expected to be applicable. Non-perturbative
methods such as lattice QCD calculations become mandatory. From these numerical simulations
the EoS of a pure gluon plasma is known to high accuracy \cite{BE96, OK99}, and there are first estimates
for the continuum EoS of systems including quarks, albeit still with unphysically large
masses \cite{FPL00}.
\pp
Various interpretations of the lattice data have been attempted in terms of physical quantities, most
prominently as the EoS of a gas of non-interacting, massive quark and gluon quasiparticles. Their
thermally generated masses are based
on perturbative calculations carried out in the hard thermal loop (HTL) scheme \cite{PKS00, PKPS96,
LU98}. More recently, the QGP has also been described in terms of a condensate of $Z_3$ Wilson lines
\cite{PI00}. Since all microscopic dynamics has been integrated out in the EoS, there exists no unique
interpretation of the lattice data, and one must resort to  additional information in order to further
restrict the
setup of such models. The phenomenological models of the QGP as an ideal gas of massive
quarks and gluons have found support from resummed perturbation theory \cite{BI01} for temperatures $T
\gsim 3 \ T_C$. However, it is astonishing that the lattice EoS is described in this model even close
to $T_C$ where one would expect non-perturbative dynamics to enter. Currently there exists not even
a qualitative microscopic explanation for this behaviour, and the model has difficulties
explaining the dropping of the thermal gluon screening mass in the vicinity of the phase
transition.
\pp
In this work we extend the quasiparticle approach. Our main new ingredient, as compared to previous work, is a
phenomenological parametrization of (de)confinement, supplemented by thermal quasiparticle masses compatible with lattice results. In
section II, we shortly review the quasiparticle formalism. Our model is
set up, and results for various
observables are compared with lattice data of continuum extrapolated SU(3) gauge theory. In section III
we give arguments how an extension to a theory including quarks might work, show how the existing
lattice
data are fitted and estimate the QCD EoS for realistic quark masses. Section IV summarizes and gives an outlook.
\section{Gluonic quasiparticles}
\subsection{Basic quasiparticle model}
In this section, we consider an $SU(N_C)$ gluon plasma ($N_C = 3)$ at finite temperature and vanishing
chemical potential. The use of a quasiparticle model in QCD is based on the observation
that in a strongly interacting system, the complex dynamics often rearranges itself in such
a way that gross features
of the physics can be described in terms of appropriate effective degrees of
freedom. From asymptotic freedom, we expect that at very high temperatures the plasma
consists of quasifree gluons. As long as the spectral function of the thermal
excitations at lower temperatures resembles qualitatively this asymptotic form, a
gluonic quasiparticle description is expected to be applicable. The
dispersion
equation for transverse
gluons reads $\omega^2 - k^2 - \Pi^*_t(\omega, k) = 0$. Here, $k = |\vec{k}|$, and
$\Pi^*_t$ is the transverse part of the thermal gluon self-energy. If, for thermal momenta
$\omega, k \sim T$, the momentum-dependence of $\Pi_t^*$ is weak and its imaginary part small, gluon quasiparticles
will propagate mainly on-shell with the dispersion relation
\be
\omega^2(k) \simeq k^2 + m^2_g(T), \label{disp_rel}
\ee
where $m_g(T)$ acts as an effective mass generated dynamically by the interaction of the
gluons with the heat bath background. Since the existence of a preferred frame of reference breaks
Lorentz invariance, new partonic excitations, longitudinal gluonic plasmons, are also present in the
plasma. However, their spectral strengths are exponentially suppressed for hard momenta and large
temperatures, so gluons are expected to retain their $\nu_g = 2(N_C^2-1)$ degrees of freedom despite
their masses.
\pp
For homogeneous systems of large volume $V$, the Helmholtz free energy $F$ is related to the pressure
$p$ by $F(T,V) = -p(T)V.$ In the present framework of a gas of quasiparticles, its explicit
expression reads
\be
p(T) = \frac{\nu_g}{6\pi^2} \int_0^\infty dk \ f_B(E_k) \frac{k^4}{E_k} - B(T), \label{p_def}
\ee
where $\nu_g$ is the gluon degeneracy factor, $E_k = \sqrt{k^2 + m_g^2(T)}$ and
\be
f_B(E_k) = \frac{1}{\exp(E_k/T) - 1}.
\ee
The energy density $\epsilon$ and the entropy density $s$ take the form
\be
\epsilon(T) = \frac{\nu_g}{2\pi^2} \int_0^\infty dk \ k^2  f_B(E_k)\ E_k + B(T) \label{e_def}
\ee
and
\be
s(T) = \frac{\nu_g}{2\pi^2 T} \int_0^\infty dk \ k^2  f_B(E_k) \ \frac{\frac{4}{3} k^2 + m_g^2(T)}{E_k}. \label{s_def}
\ee
The function $B(T)$ is introduced to act as a background field. It is necessary in order to maintain
thermodynamic consistency: eqs.(\ref{p_def}), (\ref{e_def}) and (\ref{s_def}) have to satisfy the Gibbs-Duhem relation $\epsilon + p =
sT$, and $s(T)$ is related to $p(T)$ via
\be
s = \frac{\partial p}{\partial T}. \label{s_and_p}
\ee
$B(T)$ basically compensates the additional $T$-derivatives from the temperature-dependent masses in
$p$ and is thus not an independent quantity. Since $B(T)$ adds to the energy density of the
quasiparticles in eq.(\ref{e_def}), it can be interpreted as the thermal vacuum energy
density. The entropy density, as a measure of phase space occupation, is
unaffected by $B(T)$.
\subsection{The HTL model}
In previous work \cite{PKS00, PKPS96,LU98}, the HTL perturbative expression
\be
m_g(T) = \sqrt{\frac{N_C}{6}} \ g_s(T) T \quad \mbox{with} \quad g^2_s(T) = \frac{48 \pi^2}{11N_C
\ln\left(\frac{T + T_s}{T_C/\lambda} \right)} \label{m_gluon_pert} \ee
has been used to model the thermal gluon mass\footnote{In the following, we refer to the quasiparticle model
with the gluon mass defined in eq.(\ref{m_gluon_pert}) as the HTL (hard thermal
loop) model.}. Here, $m_g(T)$ follows from the transverse part $\Pi_t^*$ of the thermal gluon
polarization tensor in the limit $k \gg g_s T$ and it is gauge-independent to this
order. Phenomenology enters in the effective coupling constant $g_s(T)$ with the two fit
parameters $T_s$ and $\lambda$. By identifying
the Landau pole of the effective coupling with a temperature close to $T_C$, the effective gluon mass
(\ref{m_gluon_pert}) becomes very heavy in the vicinity of the phase transition, and $s$, $p$ and
$\epsilon$ drop abruptly to almost zero. The resulting EoS of the HTL model is then in good
agreement with lattice data over a wide temperature range between $T_C$ and $5 \ T_C$ \cite{PKS00}. For
temperatures $T \gsim 3 \ T_C$, HTL resummed perturbation theory \cite{BI01} indeed supports the
picture of the QGP as a gas of weakly interacting, massive quasiparticles. However, there is
no {\em a
priori} reason to expect that the one-loop expressions for $g_s(T)$ and $\Pi_t^*$ can be
extrapolated to values of the
coupling as large as $g_s(T_C) \sim 2$. Furthermore, close to a phase transition, the
reliability of perturbative calculations is questionable. In the light of these facts, it is
astonishing how well the HTL model works even in the vicinity of $T_C$.
\pp
Numerical simulations suggest that the deconfinement transition in Yang-Mills theories is
second order for $N_C = 2$ \cite{order1} and weakly first order for the physical case $N_C = 3$
\cite{order2}. From the general theory of critical phenomena, it is expected that the correlation
length $\xi(T)$, which is proportional to the inverse of the gluonic screening Debye mass
$m_D$, grows when $T_C$ is approached from above (we recall that
$m_D$ measures the exponential decay of the static gluon field
correlator $\langle A^a_0(\vec{r}) A^b_0(0) \rangle_T \propto \delta^{ab}
\exp(- m_D(T) | \vec{r} | )/|\vec{r}|$). For three
colours, the mass gap does not vanish at $T_C$, so $\xi(T_C)$ remains
large, but finite. This behaviour is indeed seen in lattice calculations \cite{KK00}
: $m_D$ drops by a factor of ten when going down from $2 \ T_C$ to $T_C$ (see figure
\ref{masses}). 
\pp
In HTL perturbation theory, $m_D(T)$ and $m_g(T)$ are
connected by the simple relation:
\be
m_D = \sqrt{2} \ m_g. \label{m_d}
\ee
A scenario with heavy masses $m_g$ would then imply {\em small}
correlation lengths close to
$T_C$. It is therefore not clear how a decreasing gluonic Debye mass can
be matched to the {\em heavy, non-interacting} quasiparticles of the HTL
model. Of course, in a more general non-perturbative framework, $m_g$ and $m_D$, although both arise from the same polarization tensor, will not be related {\em a priori} by a simple constant, as in eq.(\ref{m_d}).%
\subsection{Deconfinement and the Quasiparticle model}
We believe that the failure of simple quasiparticle models to exhibit the correct temperature dependence
of the Debye mass can be traced back to the fact that the picture of a non-interacting gas
is not appropriate
close to $T_C$ because the driving force of the transition, the confinement process, is not taken into
account. Below $T_C$, the relevant degrees of freedom in a pure $SU(3)$ gauge theory are
heavy, colour singlet glueballs. Approaching $T_C$, deconfinement sets in and the gluons are liberated,
followed by
a sudden increase in entropy and energy density. Conversely, when approaching the phase transition from
above, the decrease in the thermodynamic quantities is not caused by masses becoming heavier and
heavier, instead the number of thermally active degrees of freedom is reduced due to the onset of
confinement. As $T$ comes closer to $T_C$, an increasing number of gluons gets trapped in glueballs which {\em disappear} from
the thermal spectrum:
since $m_{GB} \gsim 1.5 $ GeV and $T_C \sim 270$ MeV (for pure gauge theory), glueballs are simply too
heavy to become
thermally excited in the temperature range under consideration (up to about $5 \ T_C$). Of course,
glueball
masses may also change with temperature. However, since the contribution of glueballs to $\epsilon$,
$p$ and $s$ is negligible below $T_C$ (as evident from lattice data), this change is presumably small.
The important fact in our opinion is the following: while the confinement mechanism as
such is still not
understood, it is not necessary to know it in detail since we consider a {\em statistical}
system. All confinement does on a {\em large} scale is to cut down the number of
thermally active gluons as the
temperature is lowered. The question is whether this effect of confinement can be reconciled somehow with
the quasiparticle picture. We will show in the following that it is indeed possible in a simple, phenomenological way if we allow for an effective, temperature-dependent number of degrees of freedom $\nu_g(T)$.
\pp
Let us assume that the thermal gluon mass $m_g(T)$ does not increase as $T_C$ is approached, but instead follows roughly the behaviour of the Debye mass, {\em i.e.} it decreases. Its detailed $T$-dependence is not important for the discussion at the moment, but it will be examined in more detail in the next section. Consider now the entropy of a gas of massive gluons along eq.(\ref{s_def}) with such a dropping effective gluon mass. 
The result for $s(T)$ will clearly overshoot the lattice entropy because light masses near $T_C$ lead to an increase
in $s$. However, since the entropy is a measure for the number of active degrees of freedom, the
difference
may be accounted for by the aforementioned confinement process as it develops when the temperature
is lowered toward $T_C$. This effect can be included
in the quasiparticle picture by modifying the number of effective degrees of freedom by a temperature-dependent {\em
confinement factor $C(T)$}:
\be
\nu_g \rightarrow C(T) \ \nu_g. \label{modification}
\ee
The replacement (\ref{modification}) leads to the following expressions which replace eqs.(\ref{p_def} - \ref{s_def}):
\be
p(T) = \frac{\nu_g}{6\pi^2} \int_0^\infty dk \left[ C(T)f_B(E_k) \right] \frac{k^4}{E_k} - B(T),
\label{p2_def} \ee
\be
\epsilon(T) = \frac{\nu_g}{2\pi} \int_0^\infty dk \ k^2  \left[ C(T) f_B(E_k) \right] \ E_k + B(T)
\label{e2_def} \ee
and
\be
s(T) = \frac{\nu_g}{2\pi^2 T} \int_0^\infty dk \ k^2  \left[ C(T) f_B(E_k) \right] \
\frac{\frac{4}{3} k^2 + m_g^2(T)}{E_k}. \label{s2_def} \ee
In essence, the factor $C(T)$ represents a statistical parametrization of confinement. Its explicit functional form can be  obtained simply as the ratio of the lattice entropy and the entropy (\ref{s2_def}) calculated with a dropping input gluon mass $m_g(T)$. Qualitatively,  we expect $C(T \gg T_C) \approx 1$ at high temperatures where the deviation from the Stefan-Boltzmann limit of $\epsilon$, $s$ and $p$, as seen on the lattice, is caused solely by the thermal masses $m_g$. As the phase transition is approached from above, the number of thermally active degrees of freedom decreases and consequently, $C(T)$ becomes less than one. Finally, the entropy below $T_C$ is small, but non-zero, and we can estimate $C(T_C) \sim 0.2$ from lattice data. We would also expect that $C(T)$ is a smooth, monotonously increasing function with $T$, following the behaviour of the entropy density. 
\pp
While the lattice entropy is fitted by construction, the crucial test of the
model lies now in the reproduction of the energy density $\epsilon$ and the pressure $p$ which follow unambiguously from
eqs.(\ref{e2_def}) and (\ref{s2_def}). Owing to eq.(\ref{s_and_p}), the background field $B(T)$
depends on $m_g(T)$ and $C(T)$ through:
\bea
B(T) & = & B_1(T) + B_2(T) + B_0, \quad \mbox{where} \label{B_T} \\
B_1(T) & = & \frac{\nu_g}{6\pi^2} \int \limits_{T_C}^{T} d\tau \ \frac{dC(\tau)}{d\tau} \int
\limits_0^\infty dk \ f_B(E_k) \frac{k^4}{E_k} \quad \mbox{and} \nonumber \\
B_2(T) & = & - \frac{\nu_g}{4\pi^2} \int \limits_{T_C}^{T} d\tau \ C(\tau) \frac{d m_g^2(\tau)}{d\tau}
\int \limits_0^\infty dk \ f_B(E_k) \frac{k^2}{E_k}. \nonumber
\eea
Setting $C(T) = 1$ in $B_2$ and $B_1 = 0$, one recovers the HTL model expression for $B(T)$. The
integration constant $B_0$ is chosen such that the gluonic pressure equals the very small glueball pressure $p_{GB}$ (which is taken from the lattice) at $T_C$, according to Gibbs' condition $p_{gluon} = p_{GB}$.
\pp
Note that the proposed model for the EoS has no free fit parameters in the sense that once the non-perturbative $T$-dependence of the thermal gluon mass is fixed, $C(T)$ follows from the ratio of the entropy density of lattice QCD and the entropy density calculated with $m_g(T)$. $B(T)$ is fixed up to an integration constant, which is obtained from Gibbs' condition. Then the energy density and the pressure are uniquely determined. Since the pressure is related to the partition function $\mathcal{Z}$ by $p(T) = - F(T,V)/V = T/V \ln \mathcal{Z}$, the full thermodynamics of the system is known.
\subsection{Thermal masses}
We must now specify our input thermal quasiparticle mass $m_g(T)$ in eqs.(\ref{p2_def} - \ref{B_T}). As mentioned earlier, $m_g(T)$ is to be identified with the transverse part $\Pi_t^*(\omega, k; T)$ of the gluon polarization tensor at $\omega, k \sim T$ (see eq.(\ref{disp_rel})). Evaluating $m_g(T)$ requires a detailed non-perturbative analysis of the gluonic two-point correlation function which lattice calculations could, in principle, provide. In practice this information does not (yet) exist, so we have to rely on a model. 
\pp
Suppose we still keep the basic form of eq.(\ref{m_gluon_pert}),
\be
m_g(T) = G(T) T,
\ee
but assume that the dimensionless effective coupling $G(T)$ shows approximate critical behaviour at $T$ close to $T_C$:
\be
G(T) \simeq G_0 \left( 1 - \frac{T_C}{T} \right)^\beta, \label{G_T}
\ee
with some characteristic exponent $\beta$ and a constant $G_0$.The assumption (\ref{G_T}) implies that the thermal mass behaves as $m_g(T) \sim (T - T_C)^\beta$ close to $T_C$.  Asymptotically at $T \gg T_C$, $G(T)$ should match the HTL perturbative form as in (\ref{m_gluon_pert}). In practice we can choose this matching point, for instance, at $T_m = 3 \ T_C$. This fixes $G_0 \simeq \sqrt{N_C/6} \  g_s(T_m) \simeq 1.3$. A similar value for $T_m$ below which an explicit HTL resummation is expected to fail, was obtained in ref.\cite{AP01}.   
\pp
The quantity for which lattice information {\em does} exist is the Debye screening mass $m_D(T)$ that is related to the longitudinal part $\Pi_l^*(\omega, k; T)$ of the polarization tensor. When defined as $m_D^2 = \Pi_l^*(0, k^2 = - m_D^2)$ \cite{AR93}, the result turns out to be gauge-independent for a wide class of gauges to arbitrary order in perturbation theory \cite{KKR90} (unlike the situation for $\omega = 0$, $k \rightarrow 0$ where $\Pi_l^*$ is not gauge-invariant at next-to-leading order). While there is no {\em a priori} reason why $m_g$ and $m_D$ should still be related non-perturbatively as they are in perturbation theory ({\em cf.} eq.(\ref{m_d})), it is nevertheless instructive to recall what is known about the temperature dependence of the Debye screening mass above $T_C$. Explicit values for $m_D$ have been extracted from lattice calculations of (colour-averaged) heavy quark
potentials $V(R,T)$ \cite{KK00} by the ansatz
\be
\frac{V(R, T)}{T} \propto  \frac{ e^{-\mu(T)R} }{(RT)^d} . \label{V_pot}
\ee
Perturbation theory predicts $d=2$ and $\mu(T) = 2m_D(T)$. As elucidated in
\cite{KK00}, the potential may be better reproduced in terms of a mixture of one- and two-gluon exchange since the observed behaviour close to $T_C$
favours values of $d \sim 1.5$ in eq.(\ref{V_pot}). It is now interesting to observe that the lattice result for $\mu(T)$ can be parametrized very well by
\be
\mu(T) \simeq \mbox{const.} \cdot T \left( [1 + \delta] - \frac{T_C}{T} \right)^\beta \label{mu_T}
\ee
with $\beta \simeq 0.1$ and a small gap at $T = T_C$ introduced by $\delta \sim 10^{-6}$. The form of eq.(\ref{mu_T}) is indeed reminiscent of approximate critical behaviour. 
\pp
Let us then {\em assume} that the proportionality (\ref{m_d}) between the screening mass $m_D$ and the thermal gluon mass $m_g$ remains at least qualitatively valid in the vicinity of the phase transition, {\em i.e.} that the exponent $\beta$ in the characteristic $(1 - T_C/T)^\beta$ behaviour of both $m_D$ and $m_g$ is roughly the same. As it turns out, this is {\em not} a serious
assumption: we have checked that, as long as $m_g(T)$ and $m_D(T)$ just
have similar trends in their $T$-evolution close to $T_C$, our results are
not sensitive to the detailed quantitative behaviour of the quasiparticle
mass.
\pp
Guided by these considerations, the thermal gluon mass $m_g$ is thus parametrized as
\be
m_g(T) = G_0  T \left([1 + \delta] -  \frac{T_C}{T} \right)^{\beta}. \label{m_g}
\ee
where we allow for a small mass gap at $T = T_C$, as indicated by the lattice results for $m_D(T)$  \cite{KK00}. The small correction $\delta \ll 1$ encodes this deviation. Finally, $G_0$ is determined by the asymptotic value of the thermal mass,
 chosen such that the lattice mass and the HTL perturbative result from \cite{PKPS96} coincide at $T
\approx 3 \ T_C$, as mentioned before. In order to account for uncertainties and the  approximate nature of relation ({\ref{m_g}), we
have investigated a range of values for $G_0$, $\delta$ and $\beta$ which can be found in table
\ref{table1}. The upper limit of the
range is labelled Set A, an intermediate parameter set Set B and the lower limit Set C. Its plots and
the lattice data points for $m_D(T)$ are displayed in figure \ref{masses}.
\pp
A {\em decreasing} effective coupling strength $G(T)$ as $T_C$ is
approached from above, seems at
first sight counterintuitive: One would expect that, at a scale $T \sim \Lambda_{QCD}$,
'infrared slavery' sets in, accompanied by an {\em increasing} QCD coupling $g_s$. However, it should be
borne in mind
that this expectation is based on a perturbative result extrapolated to large couplings,
neglecting non-perturbative effects. A heuristic argument to make the dropping effective coupling plausible goes as
follows. Since we are in a strong coupling regime, the interactions between gluons cannot
be described in terms of single gluon exchange, instead they are dominated by (non-perturbative) multi-gluon dynamics. As the temperature is lowered, more and more gluons become confined and form {\em heavy}
glueballs, as outlined earlier. The effective glueball exchange interaction between gluons reduces approximately to a local
four-point interaction proportional to $1/m_{GB}^2$. The total interaction can be interpreted as a
superposition of multi-gluon and (weak) glueball exchange.
Obviously, the more glueballs are formed, the weaker becomes this interaction. The coupling $G(T)$ in eqs.(\ref{G_T}) and (\ref{m_g}) reflects an interaction between bare gluons from the heat bath on length scales $1/m_{GB}$ and larger that turns these bare gluons into massive, weakly interacting quasiparticles on length scales of order $1/T$.
%
%
%
\begin{table}[bt]
\begin{center}
\begin{tabular}{|c|c|c|c||c|c|c|}
\hline
& $G_0$ & $\delta$ & $\beta$ & $C_0$ & $\delta_c$ & $\beta_c$
\\
\hline
Set A & 1.35 & $10^{-5}$ & 0.2 & 1.24 & 0.0029 & 0.34 \\
Set B & 1.30 & $10^{-6}$ & 0.1 & 1.25 & 0.0026 & 0.31 \\
Set C & 1.30 & $10^{-7}$ & 0.05 & 1.27 & 0.0021 & 0.30 \\
\hline
\end{tabular}
\caption{Parametrizations for the thermal gluon mass $m_g(T)/T$ and the corresponding confinement
factor $C(T)$. } \label{table1}
\end{center}
\end{table}
%

%
\vspace{1cm}
\begin{figure}[htb]
\begin{center}
\epsfig{file= 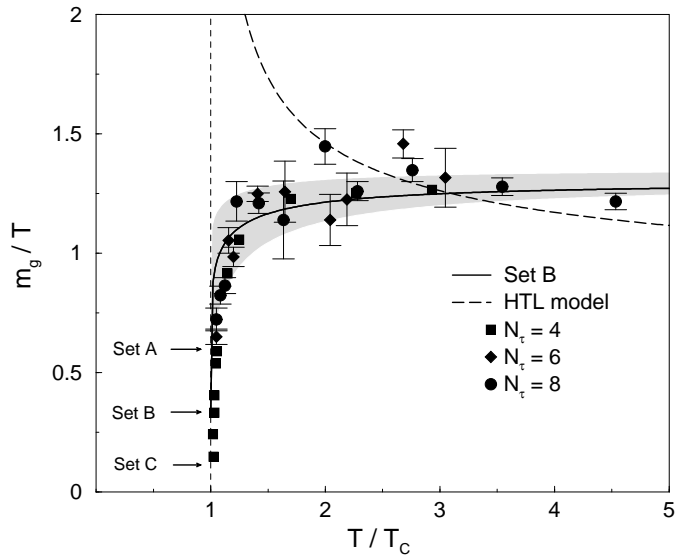, width=7.5cm, angle = -90}
\end{center}
\caption{The thermal gluon mass $m_g/T$. The grey  band shows the
parameter range of table \ref{table1} for eq.(\ref{m_g}). Set A marks the upper limit and
set C the lower, the
intermediate set B is displayed as a solid line. The arrows
indicate the mass gap $m_g(T_C)/T_C$ of the
different sets at the critical temperature. Symbols display
$\mu(T)/2$ of eq.(\ref{V_pot}), with $d = 1.5$ fixed, for lattice
configurations of
different temporal extent $N_\tau$ \cite{KK00}. For comparison the HTL
perturbative mass $m_g^{HTL}/T$ of eq.(\ref{m_gluon_pert}) is also
plotted
(dashed line),
using the parameters of ref.\cite{PKPS96}.} \label{masses}
\end{figure}
%
%
%
\subsection{Results for SU(3) gauge theory}
Now that the temperature behaviour of $m_g(T)$ is given, we can explicitly calculate the entropy density (\ref{s2_def}). Dividing the lattice entropy density by the result of this calculation, we obtain the $T$-dependence of the confinement factor $C(T)$. A very good fit of the resulting curves again exhibits an approximate critical power-law behaviour:
\be
C(T) = C_0 \left([1 + \delta_c] - \frac{T_C}{T} \right)^{\beta_c}, \label{C_paramet}
\ee
which is a non-trivial result. The corresponding parameters of $C(T)$ for the different mass
parametrizations, sets A, B and C of (\ref{m_g}), can also be found in table \ref{table1}. Their
plots are shown in figure \ref{C_T}. $C(T)$ is obviously only weakly sensitive to variations of the mass
parameters within a broad band, so the
further discussion will be based on Set B. In the following, the quasiparticle model with the gluon mass
(\ref{m_g}) and the confinement factor (\ref{C_paramet}) is referred to as 'confinement model'. For the integration constant $B_0$ appearing in eq.(\ref{B_T}) we find $0.30 \ T_C^4 \sim (200 \mbox{ MeV)}^4$. This value is about a factor of 2 larger than in the HTL model and remarkably close to the value of the bag constant at $T=0$, a welcome feature. As mentioned,  we expect the confinement effect to be negligible for $T \gsim 3 \ T_C$ where the
HTL quasiparticle model sets in. From there on, $C(T) \rightarrow 1$. The actual deviation of $C(T)$ from 1
for large temperatures has two reasons: first, even in the HTL quasiparticle model a gluon degeneracy
of $\nu_g = 16$ does not describe the data, instead a value larger by about 10\% is necessary to
account for residual sub-leading effects not captured by the model. Second, the behaviour of the gluon
mass for larger $T$ is certainly oversimplified since the parametrization $m_g \propto (T - T_C)^\beta$ is expected to be valid only in the vicinity of $T_C$. Its value overestimates the HTL perturbative result for $T \geq 3 \ T_C$ by
some 5\%, hence the thermodynamical potentials are slightly smaller than in the HTL model. One should
instead apply a smooth interpolation between the HTL perturbative mass and the $m_g(T)$ we used for
temperatures close to $T_C$. This would in turn yield a more complicated expression for $C(T)$, but it
is in principle straightforward.
\pp
%
%
\begin{figure}[tb]
\begin{center}
\epsfig{file= 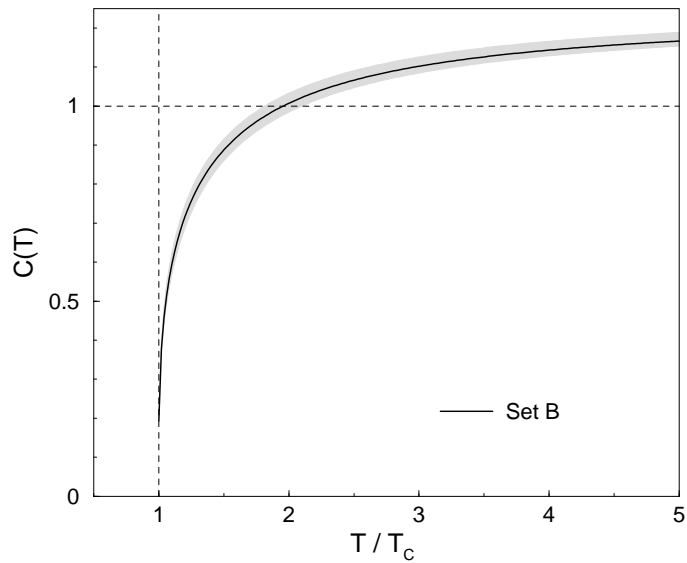, width=7.5cm, angle = -90}
\end{center}
\caption{The confinement factor $C(T)$ as a function of temperature. The grey band shows the range for
the corresponding mass parametrizations of table \ref{table1}. The solid line is obtained from set B.}
\label{C_T}
\end{figure}
%
%
%
%
\begin{figure}[tb]
\begin{center}
\epsfig{file= 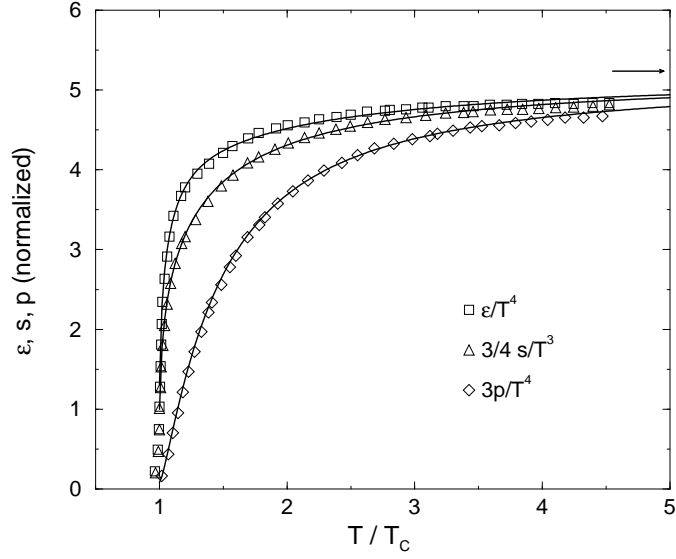, width=7.5cm, angle = -90}
\end{center}
\caption{The normalized energy density $\bar{\epsilon} = \epsilon/T^4$, entropy density $\bar{s} =
0.75 \ s/T^3$ and pressure $\bar{p} = 3p/T^4$ of our model (solid lines) compared to continuum
extrapolated SU(3) lattice data (symbols) \cite{BE96}. The size of the symbols reflects the lattice
uncertainties. The arrow indicates the ideal gas limit for massless gluons.} \label{eos_gluon}
\end{figure}
%
%
%
\begin{figure}[h!]
\begin{center}
\epsfig{file= 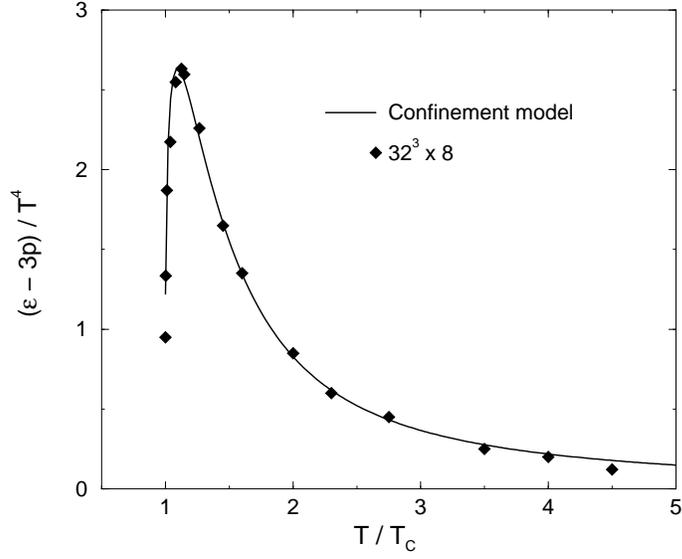, width=7.5cm, angle = -90}
\end{center}
\caption{The interaction measure $\Delta = (\epsilon - 3p)/T^4$ of the confinement model (solid line)
versus results (symbols) from a $32^3 \times 8$ lattice. The data symbols represent the continuum
interpolated values \cite{BE96}.} \label{interaction}
\end{figure}
%
%
In figure \ref{eos_gluon} we compare results of the confinement model to continuum extrapolated $SU(3)$
lattice
data \cite{BE96}. Obviously, the thermodynamic quantities are very well described even
close to
$T_C$. The slight deviations in the region $\sim 5 \ T_C$ arise from our simple parametrization of
$C(T)$. We want to stress again that the entropy density is, by construction, {\em always}
fitted. The highly
non-trivial test of
the proposed model lies in the reproduction of $\epsilon$ and $p$.
\pp
A quantity that is sensitive to the finer details of the model is the trace of the
energy-momentum tensor, $T^\mu_{\ \mu} = \epsilon -
3p$, which is compared to data from a $32^3 \times 8$ lattice in figure
\ref{interaction}. The interaction measure
\be
\Delta(T) = (\epsilon - 3p)/T^4 \label{Delta}
\ee
 is connected, via the QCD trace anomaly, to the temperature dependent gluon condensate:
\be
T^4 \Delta(T) = \langle G^2 \rangle_{T=0} - \langle G^2 \rangle_T.
\ee
Here \cite{HL93, BE96},
\be
\langle G^2 \rangle_T = \frac{11 \alpha_s}{8\pi} \langle G_{\mu\nu}^{a \ 2} \rangle_T \quad \mbox{and}
\quad \langle G^2 \rangle_{T=0} = (2.5 \pm 1.0) T_C^4.
\ee
Again, excellent agreement over the whole temperature range is observed. The confinement model
is even capable of describing the lattice data in the temperature region between $T_C$ and $1.2 \
T_C$, where the HTL model significantly underestimates the data.
\pp
%
%
%
\begin{figure}[bt]
\begin{center}
\epsfig{file= 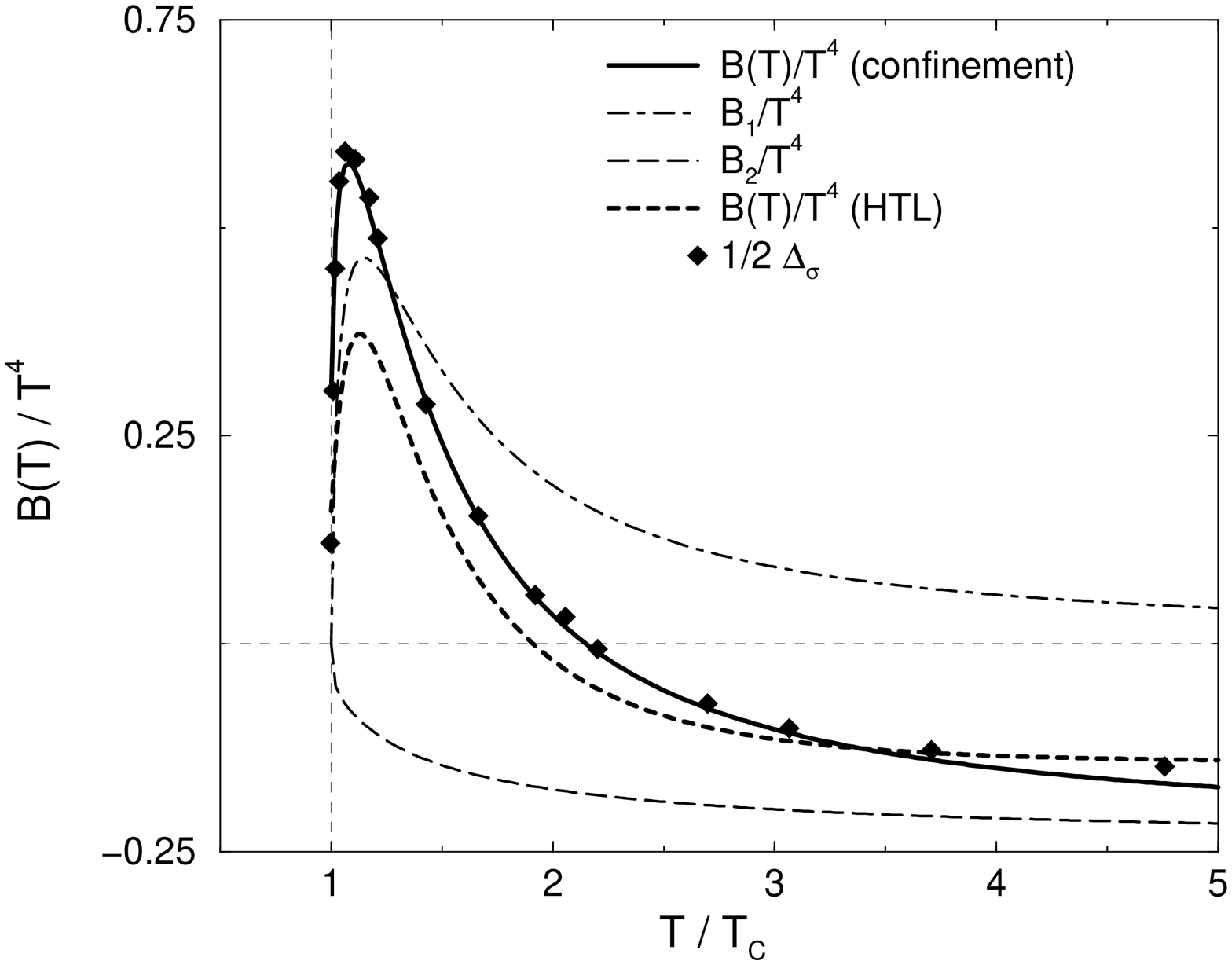, width=7.5cm, angle = -90}
\end{center}
\caption{The background field $B(T)$ and its components $B_1$ and $B_2$, defined in eq.(\ref{B_T}).
Also shown is $B(T)$ in the HTL model. Symbols display the spacelike plaquette expectation value
$\frac{1}{2}\Delta_\sigma$ taken from the lattice calculation of ref.\cite{BE96}.} \label{B_T_gluon}
\end{figure}
%
%
Finally, figure \ref{B_T_gluon} shows the background field $B(T)$ as a function of temperature.
Although the setup of the confinement model is quite different from the HTL model, the shape of this
function remains roughly the same. Note, however, that the $B_2$ term shows a completely different
temperature behaviour than in the HTL model: it is monotonously decreasing and negative from $T_C$ on.
The $B_1$ term is vital to reproduce the necessary peak structure. The maximal value of $B(T)$ is a
factor of $\sim 1.6$ larger than in the HTL model, but it also becomes negative for larger
temperatures, with its zero $T_0$ slightly shifted from $T_0 \sim 2 \ T_C$ to $T_0 \sim 2.2 \ T_C$. An
intriguing observation is that its shape now closely resembles the temperature dependence of the
spacelike plaquette expectation value $\Delta_\sigma$. The space- and timelike plaquettes, $\Delta_\sigma$ and $\Delta_\tau$,
are related to the interaction measure (\ref{Delta}) by $\Delta = \Delta_\sigma + \Delta_\tau$ and can
be expressed in terms of the thermal chromomagnetic and chromoelectric condensates $\langle
\mathbf{B}^2 \rangle_T$ and $\langle \mathbf{E}^2 \rangle_T$ as
\bea
\frac{\alpha_s}{\pi} \langle \mathbf{B}^2 \rangle_T & = & - \frac{4}{11} T^4 \Delta_\sigma +
\frac{2}{11} \langle G^2 \rangle_{T=0} \quad \mbox{and} \nonumber \\
\frac{\alpha_s}{\pi} \langle \mathbf{E}^2 \rangle_T & = & \frac{4}{11} T^4 \Delta_\tau - \frac{2}{11}
\langle G^2 \rangle_{T=0}.
\eea
What we find in fact is $B(T) = \frac{1}{2}\Delta_\sigma(T)$ (see figure \ref{B_T_gluon}). This relation between $B(T)$ and $\langle \mathbf{B}^2 \rangle_T$ may be accidental, but it may also
hint at a deeper connection between the background field as a carrier of non-perturbative effects, and
the magnetic condensate. After all, $B(T)$ represents the thermal energy of the
(non-trivial) Yang-Mills vacuum.
\pp
%
%
%
\section{Quasiparticle model with dynamical quarks}
%
%
%
The extension of the mechanism presented in the last section to systems with dynamical
quarks is not straightforward. Simulations of fermions on the lattice are still plagued by problems. No
concise continuum extrapolation of the QCD EoS with realistic quark masses exists to date.
Nevertheless, it is still possible to construct a model of the EoS with quarks, using some reasonable
arguments based on the available lattice data.
\subsection{Lattice QCD thermodynamics with quarks}
There have been lattice calculations of the pressure with different numbers of quark flavours $N_f$
\cite{FPL00, AK01_2}. In the following we focus on results of the Bielefeld group in ref.\cite{FPL00}
where a p4-improved staggered action on a $16^3 \times 4$ lattice was used. There, the
$N_\tau$-dependence is known to be small, in contrast to the standard staggered fermion actions which
show substantially larger cut-off effects. Lattice calculations were performed for two and three
flavour QCD with quarks of mass $m_q/T = 0.4$, and for three flavours with two light quarks ($m_q/T =
0.4$) and an additional heavier quark ($m_s/T = 1.0$). From the experience in the pure gauge sector, it
has
been estimated that the continuum EoS lies about 10 - 20\% above the data computed on finite lattices.
\pp
Figure \ref{pressure_scaled} displays the lattice pressure, normalized to the Stefan-Boltzmann ideal
gas value, for the pure gauge system and for systems with 2, 2+1 and 3 quark flavours. A striking
feature is
that, within the errors arising from the cut-off dependence, the QCD EoS shows a remarkable flavour
{\em independence} when plotted against $T/T_C$. This picture suggests that the flavour dependence is well approximated by a term
reminiscent of an ideal gas,
\be
p(T, N_f) \propto \left(16 + \frac{7}{4} \cdot 2N_C N_f \right) \frac{\pi^2}{90} \
\tilde{p}(T/T_C) \label{p_tilde}
\ee
with a function $\tilde{p}(T/T_C)$. Since $T_C$ changes obviously with the number of degrees of freedom present in the thermal system (and therefore with $N_f$), $\tilde{p}$ is also implicitly $N_f$-dependent. Scaling against $T/T_C$, however, the shape of $\tilde{p}$ remains almost the same, indicating that the confinement mechanism itself is only weakly flavour-dependent. Therefore, once we understand the machinery that
is responsible for $\tilde{p}(T)$ in the gluon sector, an extrapolation to systems including quarks appears
feasible, with the confinement model suitably adapted. Lattice results on the order of the phase transition in full QCD support this idea
by indicating that the transition is first order in the case of three light, degenerate quark flavours
and most likely second order for two flavours \cite{AK01, CB00, YI96}. Note however that,
after applying the previously mentioned 10-20\%  correction to the lattice data, the
continuum
estimate of the pressure with dynamical quarks is much closer to the ideal gas limit than in the pure
gauge sector.
%

\begin{figure}[h!]
\begin{center}
\epsfig{file= 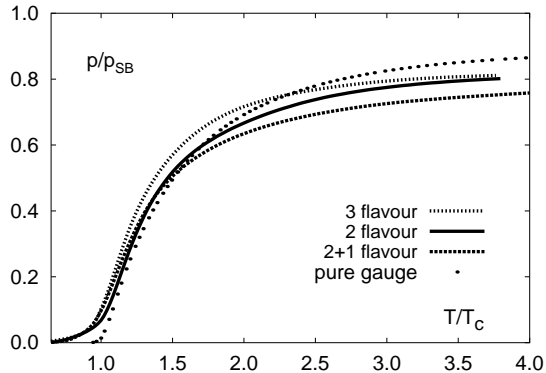, width=7.5cm}
\end{center}
\caption{The  pressure, normalized to the Stefan-Boltzmann ideal gas value, for the
continuum-extrapolated pure gauge system and for systems with 2, 2+1 and 3 flavours on a $16^3 \times
4$ lattice, obtained with a p4-improved staggered fermion action. The continuum limit is estimated to
lie about 10 - 20\% above the curves shown (figure from \cite{FK01}).} \label{pressure_scaled}
\end{figure}


%
\subsection{Thermal masses}
No lattice data on thermal masses with dynamical quarks are available. We thus construct effective
masses for quarks and gluons by assuming that the $N_C$- and $N_f$-dependence of $m_g$ and $m_q$ are both given by the HTL asymptotic limit. For the thermal gluon mass we employ the ansatz:
\be
\frac{m_g(T)}{T} = \sqrt{\frac{N_C}{6} + \frac{N_f}{12}} \ \tilde{g}(T, N_C, N_f) \label{m_g_2}
\ee
with the effective coupling
\be
\tilde{g}(T, N_C, N_f) = \frac{g_0}{\sqrt{11N_C - 2N_f}} \left( [1 + \delta] - \frac{T_C}{T}
\right)^{\beta}. \label{g_tilde}
\ee
$g_0$, $\delta$ and $\beta$ are taken to be universal. Setting $g_0 = 9.4$, $\delta = 10^{-6}$ and
$\beta = 0.1$, the gluon and quark masses coincide with the two flavour HTL masses at $T \simeq 3 \
T_C$,
using the parameters of ref.\cite{PKS00}. The thermal quark mass becomes \cite{PKS00}
\be
\frac{m_q(T)}{T} = \sqrt{ \left( \frac{m_{q,0}}{T} + \sqrt{\frac{N_C^2-1}{16 N_C}} \tilde{g}(T)
\right)^2 + \frac{N_C^2-1}{16 N_C} \tilde{g}(T)^2  } \label{m_quark}
\ee
with the zero-temperature bare quark mass $m_{q,0}$. 
\pp
In \cite{ST99}, a non-perturbative dispersion equation for a thermal quark interacting with the gluon condensate has been calculated, and it has been
found that the effective quark mass is given by $m_q \simeq 1.15 \ T$ in the temperature range between $1.1
\ T_C$ and $4 \ T_C$. Eq.(\ref{m_quark}) is within 10\% in agreement with this result. Nevertheless,
very close to $T_C$ the parametrization (\ref{m_quark}) may be too simple: if the expected chiral phase
transition is second order (or weakly first order), fermions may decouple in the vicinity of the phase
transition because they have no Matsubara zero modes, and the transition dynamics would be dominated by
the bosonic gluons only. In this case gluon masses should become independent of $N_f$. However, as in the
pure gluon sector, the results are stable against small variations of the mass parametrizations, and as
long as no further information is available, eqs.(\ref{m_quark}) and(\ref{m_g_2}) may be taken as an
educated guess.
\pp
%
%
\begin{table}[bt]
\begin{center}
\begin{tabular}{|c|c|c|c|}
\hline
& $C_0$ & $\delta_c$ & $\beta_c$ \\
\hline
2 flavours & 1.25 & 0.02 & 0.28 \\
2+1 flavours & 1.16 & 0.02 & 0.29 \\
3 flavours & 1.03 & 0.02 & 0.2 \\
\hline
gluon & 1.25 & 0.0026 & 0.31 \\
\hline
\end{tabular}
\caption{Parametrizations of eq.(\ref{C_paramet}) for the confinement function $C(T)$ in the presence
of dynamical quark flavours. For comparison, the corresponding values of the pure gauge system (set B) are also
shown.} \label{table2}
\end{center}
\end{table}
%
%
We proceed now as follows: First, we assume that the continuum limit of the pressure
 can be obtained from the $N_\tau = 4$ lattice data by applying a 10\% correction as mentioned, {\em
i.e.} $p_{cont} \simeq 1.1 \ p_{lat}$. Second, using Occam's razor we employ a {\em universal} confinement function $C(T)$
for both quarks and gluons, motivated by eq.(\ref{p_tilde}).  The extension of eq.(\ref{p2_def}) to systems including quark flavours is straightforward:
\be
p(T) = \frac{\nu_g}{6\pi^2} \int_0^\infty dk \left[ C(T)f_B(E_k^g) \right] \frac{k^4}{E_k^g} + \sum
\limits_{i=1}^{N_f}  \frac{2 N_C}{3\pi^2} \int_0^\infty dk \left[ C(T)f_D(E_k^i) \right]
\frac{k^4}{E_k^i} - B(T). \label{p3_def}
\ee
Here $E_k^g = \sqrt{k^2 + m_g^2(T)}$ as before, and $f_D(E) = (\exp(E/T) + 1)^{-1}$. The quark energy is $E_k^i = \sqrt{k^2 + m_i^2}$ for each quark flavour $q = i$, and $m_q(T)$ is given by eq.(\ref{m_quark}) with the bare quark masses $m_{q,0}$. The background field $B(T)$, the entropy density $s$ and the energy density $\epsilon$
follow analogously. If the confinement model is applicable, we should expect that the parameters of
$C(T)$ in eq.(\ref{C_paramet}), as shown in table \ref{table1} for the gluonic calculations, are
similar in the presence of quarks. We start therefore with the gluon values for $C_0$, $\delta_c$ and
$\beta_c$ and vary them slightly until good agreement with the lattice pressure is obtained. To account
for the temperature-dependent {\em bare} masses used specifically in the lattice calculations, the
quark masses $m_{q,0}$ in eq.(\ref{m_quark}) are set to $m_{q,0} = 0.4 \ T$ (light quarks) and $m_{s,0}
= 1.0 \ T$ (heavy quark). Figure \ref{pressure_quarks} shows the results for 2 and 2+1 flavours, the
corresponding values for the parameters of $C(T)$ can be found in table \ref{table2}. We observe that
indeed, the confinement factor $C(T)$ does not differ much from the pure gluonic case. The factor $B_0$
is set to $1.4 \ T_C^4 \simeq$ (180 MeV)$^4$. The larger value for $\delta_c$ is explained by noting
that, for $T < T_C$, many light quark-antiquark composites (pions, kaons etc.) are present. They
contribute sizably to the entropy in the hadronic phase. Accordingly, $C(T_C)$ is larger than in the
pure gluon case, hence $\delta_c$ has to increase.
%


\begin{figure}[tbh]
\begin{center}
\epsfig{file= 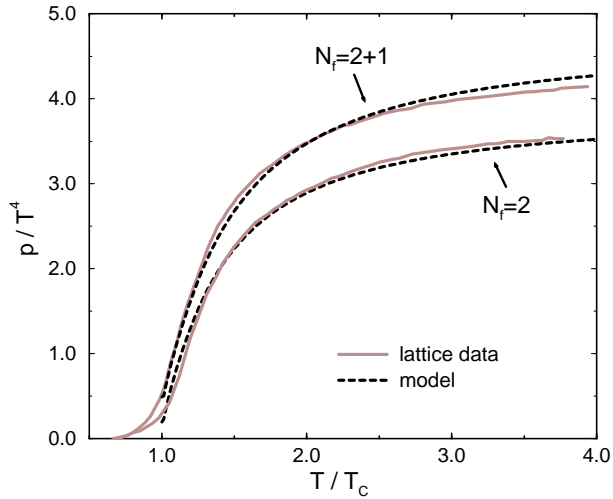, width=7.5cm, angle = -90}
\end{center}
\caption{The rescaled lattice pressure $p_{cont} \simeq 1.1 p_{lat}$ (grey lines) for 2 and 2+1
flavours and the pressure obtained from the confinement quasiparticle model with running bare quark
masses (dashed lines). Values for the parameters of $C(T)$ are shown in table \ref{table2}.}
\label{pressure_quarks}
\end{figure}


\subsection{Physical quark masses}
%
%

\begin{figure}[tbh]
\begin{center}
\epsfig{file= 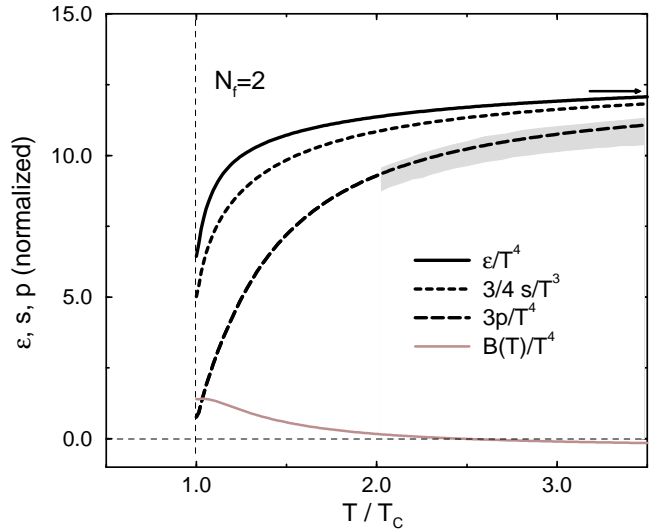, width=8.0cm, angle = -90}
\end{center}
\caption{Pressure, energy and entropy density for two light quark flavours in the confinement model.
The arrow indicates the ideal gas limit. The grey band is an estimate of the continuum EoS for
massless two flavour QCD, based on an extrapolation of lattice results \cite{FPL00}.} \label{2_flav}
\end{figure}

%
%

\begin{figure}[bth]
\begin{center}
\epsfig{file= 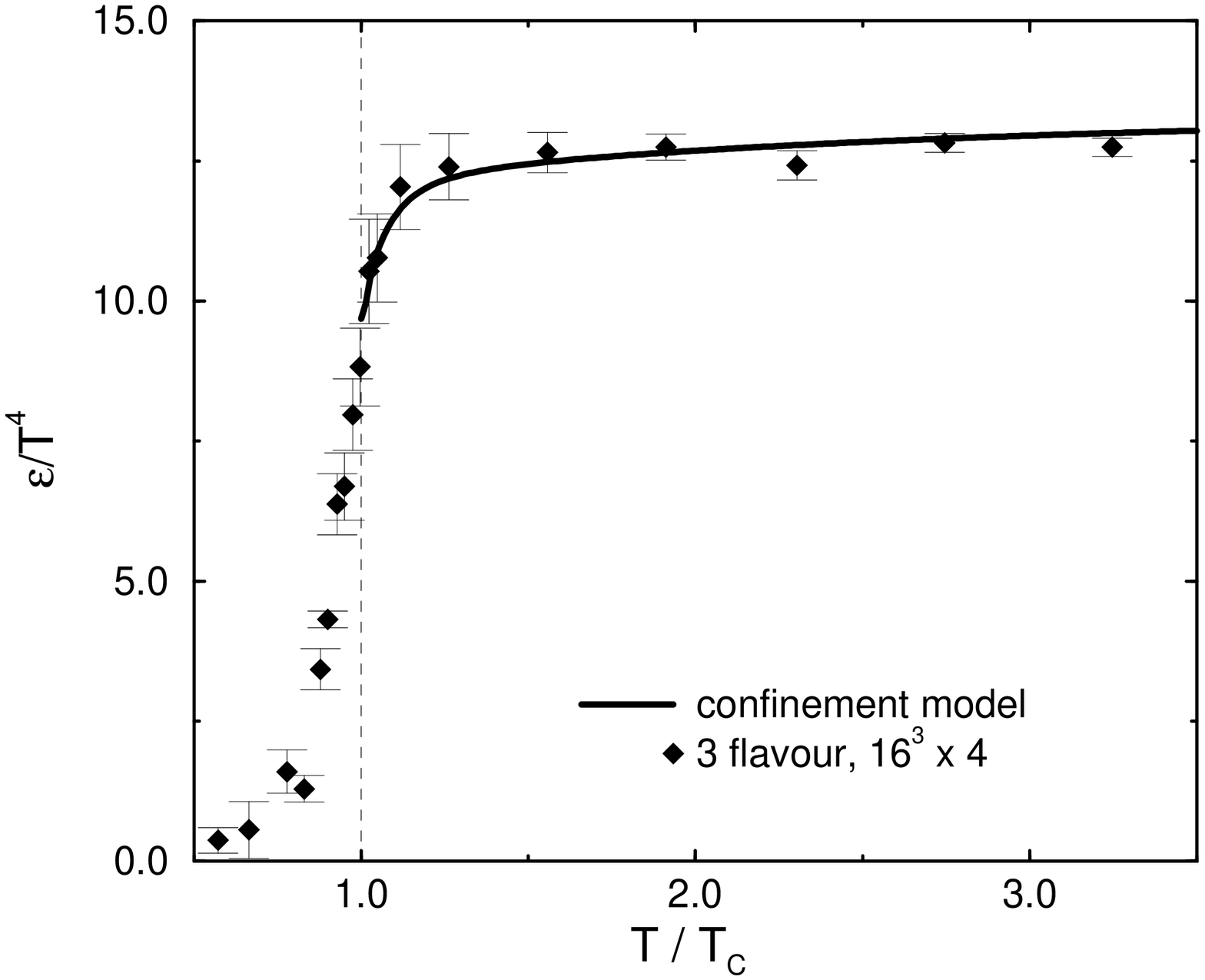, width=8.0cm, angle = -90}
\end{center}
\caption{Energy density estimate in the chiral limit for three quark flavours on a $16^3 \times 4$ lattice \cite{FK01} (data points) and as obtained in the confinement model (solid line).} \label{3_flav}
\end{figure}

%
Pressure, energy and entropy density for physical quark masses are finally obtained by setting
$m_{q,0}$ in eq.(\ref{m_quark}) to the real-world values $m_{u,d} \simeq 0$ and $m_s \simeq 170$ MeV.
This procedure assumes that $C(T)$ is independent of $m_{q,0}$ which is not clear. In present lattice
simulations, the pions are too heavy, $m_\pi^{lat} \geq $ 450 MeV, therefore their contribution to $s$
or $p$ is strongly Boltzmann suppressed. Since $$\frac{e^{-m_\pi^{lat}/T_C}}{ e^{-m_\pi^{phys}/T_C}}
\approx \frac{1}{7},$$ future computations with lighter, more realistic pion masses are expected to
find a substantially larger pressure and entropy in the hadronic phase below $T_C$. With this {\em caveat}, we assume for now
that $C(T)$ does not depend on $m_{q,0}$. Figure \ref{2_flav} shows a prediction of $\epsilon$, $s$ and
$p$ for massless two flavour QCD. Reassuringly, the pressure of the confinement model is well within
the estimate for the continuum EoS of ref.\cite{FPL00} for $T > 2 \ T_C$. In contrast to the pure gluon
EoS, we observe that the energy and entropy are close to the ideal gas limit already at $T = 3 \ T_C$.
However, it has to be borne in mind that their normalization is set by $C_0$ which in turn depends on
the continuum estimate of the $N_\tau = 4$ lattice data. More reliable estimates for the continuum
pressure are needed to confirm this behaviour. It is also worthwhile noting that, going from
temperature-dependent bare masses (as used in the lattice simulations) to the chiral limit, the
corresponding change of the pressure in the confinement model is stronger than expected from an ideal
Fermi gas. It rises by about 7\% whereas for an ideal gas with quark mass $m_q/T = 0.4$ the difference
would be only about 3.5\% (for $N_f = 2$).
\pp
In figure \ref{3_flav}, we plot the energy density for three light quark flavours on a $16^3 \times 4$
lattice \cite{FK01} and as obtained in the confinement model. Since there is no estimate of the
continuum limit of these lattice data, the normalization, set by $C_0$ in eq.(\ref{C_paramet}), is
substantially smaller than in previous cases. Apart from that, the data are very well reproduced down to
$T_C$. Finally, figure \ref{2_1_flav} shows the calculated pressure, energy and entropy density for a system with
two light quark flavours ($m_{q,0} = 0$) and a heavier strange quark ($m_{s,0} \simeq 170 $ MeV). Here, the approach to the
Stefan-Boltzmann limit is obviously slower than in the two flavour case because of the mass suppression
of the third, heavier flavour. $N_f$ in eqs.(\ref{m_g_2}), (\ref{g_tilde}) and (\ref{m_quark}) was set
to 2.3. The results for $\epsilon$ and $p$ are also in good agreement for $T > 2 \ T_C$ with an EoS
obtained in the HTL quasiparticle model \cite{AP98}. Closer to $T_C$, the confinement model predicts a
stronger decrease of the energy density, though. Again, the background field $B(T)$ resembles the shape
of the corresponding function in the HTL model. However its zero, $T_0$, is considerably shifted, from
$T_0 \sim 1.7 \ T_C$ to $T_0  \sim 2.7 \ T_C$.
%
%
\begin{figure}[h!]
\begin{center}
\epsfig{file= 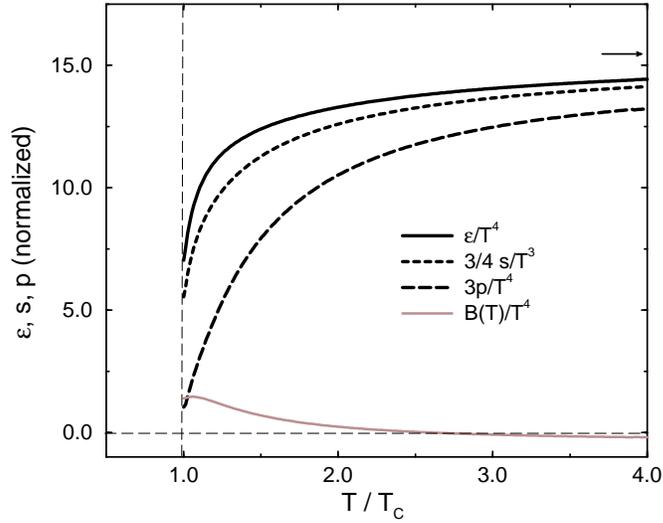, width=8.0cm, angle = -90}
\end{center}
\caption{Pressure, energy and entropy density for two light quark flavours ($m_{u,d} = 0$) and a heavier strange quark
($m_s \simeq 170 $ MeV) in the confinement model. The arrow indicates the ideal gas limit of massless
three-flavour QCD.} \label{2_1_flav}
\end{figure}
%
%
%

\section{Summary and outlook}
We have presented a novel quasiparticle description of the QCD EoS at
finite temperature. Our main
modification as compared to previous work is the schematic inclusion of confinement thas has so far been neglected. First,
we have considered thermal SU(3) gauge theory. Guided by lattice results for the thermal screening masses, we constructed a thermal mass $m_g(T)$ for the gluon and calculated the
corresponding entropy density as a measure of phase space occupation. The
difference to the lattice entropy was then attributed to the confinement process setting in  as $T_C$ is
approached from above. Since confinement simply reduces the number of thermally active degrees of
freedom in a
statistical sense, this non-perturbative behaviour is incorporated in a
model of quasifree, massive
quasiparticles by a modification of the particle distribution functions with a confinement factor
$C(T)$. The energy density and the pressure are then uniquely determined. Our model agrees very well
with
continuum-extrapolated lattice data. The interaction measure $\Delta(T)$ is even better reproduced than
in
the HTL model. As a new and interesting aspect, a possible physical
connection of the background field $B(T)$
(which is the thermal energy of the Yang-Mills vacuum) with the chromomagnetic condensate
$\langle \mathbf{B}^2 \rangle_T$ has been proposed. Our results turn out
not to be very sensitive to the detailed parametrization of the $T$-dependent quasiparticle
mass; the driving new feature is the confinement factor $C(T)$.
\pp
For systems with dynamical quarks, sufficiently precise lattice data are not
yet available. With reasonable assumptions, we can nevertheless reproduce
existing continuum estimates
of
lattice results for the pressure. Of course, future simulations with higher statistics and smaller quark
masses are needed to judge if our confinement model yields sensible results in the presence of quark
flavours. It is encouraging, though, that for temperatures $T \gta 2 \ T_C$, the predictions
of the confinement model with realistic  quark masses agree well with HTL model calculations and an
estimate of the continuum EoS in the chiral limit.
\pp
The function $C(T)$ parametrizes our ignorance about details of the confinement
mechanism.  It would be
desirable to connect this macroscopic quantity with microscopic, first-principle QCD dynamics,
preferably starting in the gluon sector. Interesting questions arise in this context, as to the r\^{o}le
of instantons and condensates of magnetic monopoles.
We point out that a comparison of $B(T)$ with lattice data for the spacelike plaquette $\Delta_\sigma$ in the  presence of
quarks may shed more light on such conjectures.
\section{Acknowledgements}
We thank Norbert Ligterink and Thorsten Renk for stimulating discussions. We are particularly grateful to  Ulrich Heinz for his very thorough and careful reading of the manuscript and for his comments.

\end{document}